\documentclass[letterpaper]{article}
\usepackage{isea}
\usepackage[pdftex]{graphicx}
\usepackage{times}
\usepackage{helvet}
\usepackage{courier}
\usepackage[numbers]{natbib}
\title{She Works, He Works: Critical Explorations of Gender in AI-Generated Representations}
\author{Amalia Foka\\
School of Fine Arts, University of Ioannina\\
Ioannina, Greece\\
afoka@uoi.gr\\
\newline
\newline
}
\begin{document}
\maketitle
\begin{center}\footnotesize\emph{Author's version of the paper published in the Proceedings of the 30th International Symposium on Electronic Art (ISEA 2025), Seoul, Republic of Korea, 23--29 May 2025. This version supersedes the earlier preprint ``She Works, He Works: A Curious Exploration of Gender Bias in AI-Generated Imagery''. \copyright~2025 Amalia Foka.}\end{center}
\begin{abstract}
\emph{She Works, He Works} is an interdisciplinary project that combines feminist visual theory, AI critique, and artistic practice to examine gender representation in AI-generated imagery. By juxtaposing text-to-image outputs of male and female construction workers, it reveals how societal biases manifest in generative AI systems. The study employs critically crafted prompts that blend traditionally gendered traits with unconventional contexts, exposing disparities in representation-emphasizing aesthetics and agility for women versus authority and professionalism for men. Grounded in Griselda Pollock's feminist insights on visual culture, the project positions AI as both a technological and cultural actor, capable of perpetuating stereotypes and enabling critique. As both a study and an artwork, the project underscores the transformative potential of AI to reimagine gender norms and inspire equitable narratives. It highlights the ethical implications of AI in creative industries, calling for transparent practices, diverse datasets, and reflexive engagement with AI's cultural impact. Through its integration of critical analysis and artistic exploration, \emph{She Works, He Works} contributes to broader discussions on inclusive AI development and the role of art in fostering societal change. This dual approach showcases the power of AI as a medium for critique and a tool for envisioning more just and inclusive futures.

\end{abstract}

\keywords{Keywords}

Gender representation, text-to-image models, critical prompt engineering, AI-generated art, bias in AI, feminist visual theory, cultural representation, ethical AI development, societal stereotypes

\section{Introduction}
The emergence of sophisticated text-to-image AI models, such as DALL-E and Midjourney, has catalyzed an intense exploration of their creative potential in the art world. These models offer unprecedented capabilities for generating visually compelling content from textual descriptions, sparking innovative applications in art, design, and communication. However, alongside their impressive functionalities, critical concerns have emerged regarding the societal biases embedded within these systems. Early investigations reveal that such biases often manifest in the representation of gender, perpetuating stereotypes and reinforcing existing power imbalances. These issues underline the urgency of addressing the ethical dimensions of generative AI, particularly in contexts that influence cultural perceptions.

\emph{She Works, He Works} is an interdisciplinary project that merges artistic practice with critical inquiry to examine how text-to-image AI models represent gender in professional contexts. Focusing on the construction industry -a field historically dominated by men- this project generates and juxtaposes AI-created diptychs depicting male and female construction workers. Using identical prompts differing only by gender pronouns, the work interrogates the biases inherent in these systems. The resulting images not only serve as data for analysis but also as curated visual artifacts, prompting viewers to reflect on the disparities in gender representation. This dual framing positions the audience as active participants in a broader cultural critique.

This project draws on feminist visual theory, particularly Griselda Pollock's exploration of sexual difference and the politics of representation \cite{pollock_vision_difference}, situating AI systems as contemporary tools of visual representation. These systems, like traditional art forms, reflect societal norms and power structures, yet they also hold the potential to disrupt and reimagine them. By focusing on the construction industry, the project explores how generative AI navigates gendered visual tropes in a context that symbolizes strength, authority, and skill-qualities historically coded as masculine.

The thesis of this project is twofold: as an artistic exploration, \emph{She Works, He Works} interrogates the biases encoded in AI-generated imagery, and as a critical study, it seeks to reveal how these biases perpetuate or challenge existing power dynamics. By integrating the methodologies of art, AI critique, and feminist theory, the project situates itself within the broader discourse on ethical AI development and cultural representation.

This work's relevance extends beyond academic critique. As generative AI becomes more prevalent in shaping media and culture, it is imperative to explore its impact on societal norms and creative expression. \emph{She Works, He Works} contributes to this discourse by demonstrating how critical and artistic methodologies can intersect to foster inclusivity and reflexivity in AI technologies. Through its artistic and analytical lens, the project positions generative AI as both a mirror of societal biases and a tool for imagining more equitable futures, inviting audiences to engage with its transformative potential.

\section{Background and Related Work}
The representation of gender in visual culture has been a pivotal focus of feminist scholarship, providing a foundational lens for critically examining biases in AI-generated art. Griselda Pollock's seminal work, \emph{Vision and Difference} \cite{pollock_vision_difference}, offers a framework for understanding how visual representations of women are interwoven with societal power dynamics and gender ideologies. Pollock posits that these representations often function as "signs" that reinforce traditional gender roles, restricting female agency. This theoretical grounding inspires scrutiny of how generative AI technologies perpetuate or challenge these entrenched inequalities in contemporary visual culture.

Expanding on Pollock's insights, feminist scholars have analyzed visual culture through intersectional lenses. Amelia Jones' \emph{The Feminism and Visual Culture Reader} \cite{jones_feminism_visual_reader} and Victoria Horne and Lara Perry's \emph{Feminism and Art History Now} \cite{horne_feminism_art} emphasize inclusivity and the intersection of race, class, and gender in artistic production and reception. Laura Mulvey's concept of the "male gaze" \cite{mulvey_visual_pleasure} remains essential for dissecting objectification and sexualization in media, while Erving Goffman's \emph{Gender Advertisements} \cite{goffman_gender_advertisements} and subsequent critiques \cite{obarr_advertising} explore how stereotypes evolve across platforms. Scholars such as Nancy Wang Yuen \cite{yuen_reel_inequality} and Gillian Rose \cite{rose_visual_methodologies} further contextualize gendered representations within racial and class-based hierarchies, emphasizing the layered nature of bias in visual narratives.

Generative AI technologies introduce novel complexities to this discourse. By relying on vast datasets often curated without socio-historical context, AI systems inherit and amplify biases rooted in their training materials. For instance, Bolukbasi et al. \cite{Bolukbasi2016} reveal how gender stereotypes are embedded in word embeddings, while Buolamwini and Gebru's \emph{Gender Shades} \cite{Buolamwini2018} demonstrate racial and gender disparities in facial recognition systems. Cheong et al. \cite{Cheong2024} and Birhane et al. \cite{birhane2021} expose the ways in which text-to-image models reinforce stereotypes across gender, race, and class, raising concerns about the societal impact of these technologies.

Theoretical critiques of generative AI further illuminate its role in cultural production. Crawford and Paglen's \emph{Excavating AI} \cite{Crawford2021ExcavatingAT} critiques the socio-political implications of dataset curation, exposing how AI normalizes and encodes inequities. Ruha Benjamin's "New Jim Code" \cite{benjamin_race_after_technology} explores how AI systems deepen systemic discrimination under the guise of neutrality. Mimi Onuoha's \emph{The Library of Missing Datasets} \cite{onuoha_library_missing_datasets} highlights the societal impact of omissions in data-driven systems, underscoring the underrepresentation of marginalized communities. Joanna Zylinska's \emph{AI Art: Machine Visions and Warped Dreams} \cite{zylinska_aiart} critiques the commodification of creativity in generative AI and its embedded labor dynamics, offering critical insights into the ethics of AI-mediated artistic production.

Intersectionality plays a critical role in understanding these biases. Studies of visual culture frequently demonstrate how gender, race, and class intersect to produce layered inequities. For example, Yuen's \emph{Reel Inequality} \cite{yuen_reel_inequality} explores how Hollywood perpetuates racialized and gendered stereotypes, providing a lens to interrogate how AI reflects similar patterns. Posthumanist critiques further complicate these dynamics by examining AI as both a product and a producer of cultural meaning, shaping how identities are constructed and perceived in the digital age \cite{braidotti_posthuman}.

Artists and theorists have responded to these challenges through "creative misuse," employing generative technologies in ways that subvert their intended functionalities. Eryk Salvaggio's \emph{Flowers Blooming Backward Into Noise} \cite{salvaggio_noise_feedback} uses diffusion models to critique stereotypes, while \emph{Because of You} examines the abstraction and exploitation of human data through historical parallels \cite{because_of_you}. Caroline Sinders critiques emotion recognition systems for their cultural biases \cite{sinders_anger_disgust_2019}, and Supercomposite's \emph{Loab} highlights the unpredictability and ethical dilemmas inherent in generative AI \cite{swanson_loab}. Trevor Paglen's \emph{ImageNet Roulette} \cite{paglen_imagenet_roulette} exposes biases in computer vision datasets, inviting audiences to confront the problematic classifications embedded in these systems.

\emph{She Works, He Works} situates itself within these academic, artistic, and theoretical discourses. By combining Pollock's framework with the critiques of Benjamin, Onuoha, and Zylinska, the project interrogates the biases encoded in AI-generated visual content. Through critical prompt engineering, it examines how text-to-image models depict gender in professional contexts, using the construction industry as a lens to explore intersections of power, labor, and representation.

This project leverages artistic methodologies to engage critically with AI, positioning its outputs as both evidence of societal biases and artifacts for cultural critique. By integrating feminist visual theory, intersectionality, and creative practice, \emph{She Works, He Works} contributes to broader efforts to develop equitable and inclusive AI technologies. It highlights the transformative potential of art and critique in reimagining AI's role in shaping more just cultural narratives.

\section{Methodology}
This project employs a process of critical prompt engineering, drawing upon feminist theories of visual culture and contemporary research on biases in AI systems, to examine the gendered representations produced by text-to-image models. The methodology integrates principles of "creative misuse" to subvert and interrogate the assumptions embedded within generative AI systems. By intentionally designing prompts to juxtapose traditionally gendered language with unexpected scenarios, the approach aims to reveal how AI systems trained on potentially biased data construct and perpetuate societal stereotypes.

The process of crafting prompts began with an analytical framework designed to explore how AI systems reconcile descriptive elements traditionally associated with masculinity or femininity. Each prompt was constructed to challenge the AI's implicit biases by blending attributes associated with leadership, physicality, and expertise (often coded as masculine) with language evoking artistry, precision, or emotional resonance (often coded as feminine). This combination sought to disrupt the default patterns of representation embedded in the models and encourage nuanced outputs.

While crafting the prompts, efforts were made to avoid reinforcing stereotypes through the language choices themselves. Descriptive terms traditionally coded as masculine or feminine were deliberately paired to balance their representation. Additionally, metaphorical and poetic elements were integrated to stretch the interpretive capacity of the AI systems and elicit outputs that reveal underlying biases. By using identical prompts, differing only by gendered pronouns, the methodology ensures that discrepancies in output stem from the AI's interpretive processes rather than prompt biases. This approach, while not eliminating all biases, seeks to mitigate their impact by maintaining fairness and consistency in the prompt design.

The construction industry was chosen as a focal point for this study because of its strong cultural and visual associations with masculinity. This field provides a potent lens for examining how AI systems interpret and reproduce gender norms in a context that challenges traditional representations of women. Additionally, construction work combines physicality, leadership, and expertise, offering rich opportunities to explore how AI balances these attributes across genders. Future studies might extend this analysis to other professions, such as caregiving or technology, to reveal whether AI systems exhibit similar patterns of bias in different social and professional contexts.

\subsection{Prompts and Creative Misuse}
Each prompt was intentionally crafted with poetic and metaphorical language to test the models' interpretive limits and encourage nuanced outputs. The inclusion of detailed descriptions and complex metaphors was a deliberate methodological choice to simulate richer narrative contexts and probe how these systems navigate the linguistic ambiguities and contrasts that characterize human expression. Below are the detailed examples of prompts and the application of creative misuse:
\begin{itemize}
        
    \item \textbf{Dawn's Symphony of Industry}: This prompt introduces a "symphony of steel and concrete" and describes "movements graceful yet purposeful, like dancers in a ballet of construction." By combining industrial imagery traditionally coded as masculine with artistic descriptors such as "symphony" and "ballet," typically associated with femininity, the prompt challenges the AI to reconcile these contrasting attributes. The "creative misuse" in this context lies in exploring how the models respond to leadership qualities when described with both feminine- and masculine-coded language, using gender-specific pronouns to analyze variations in depictions of male and female site managers.
        
    \item \textbf{A Tapestry of Skill and Grit}: This prompt describes the site manager as "weaving" a "tapestry," evoking the traditionally feminine craft of weaving, set against the rugged backdrop of construction. It blends concepts of meticulous skill and perseverance to explore how the AI balances these attributes in its representation of gendered expertise. Additionally, the prompt tests how the models interpret the actions of "guiding workers" with a "voice firm yet encouraging, ensuring each task is executed flawlessly," and the manager's "watchful eye" for both the construction and the workers. The "creative misuse" examines whether the AI distinguishes male and female site managers in the portrayal of leadership and attentiveness, revealing potential biases in how authority and expertise are gendered.
        
    \item \textbf{A Conductor of Controlled Chaos}: The metaphor of a conductor leading a symphony frames the site manager's leadership in terms of harmony and precision, contrasting with the chaotic, noisy environment of construction. This prompt probes whether the AI associates female authority with collaborative orchestration rather than directive or authoritarian styles. Additionally, the prompt explores how the models interpret descriptions of the manager's "keen eye spotting potential hazards" and "quick thinking averting disaster." These phrases highlight vigilance and decision-making under pressure, qualities often coded as masculine. The "creative misuse" lies in examining whether the AI portrays male and female site managers differently when performing critical, high-stakes tasks, and whether it assigns distinct visual or narrative attributes to their depictions of authority and expertise.
        
\item \textbf{A Symphony of Sweat and Sacrifice}: By emphasizing "sweat" and "sacrifice," this prompt highlights resilience and emotional investment in a challenging environment. The imagery of a rain-soaked, muddy construction site contrasts with the triumphant structure emerging from the chaos, encouraging the AI to portray female determination without resorting to stereotypes of fragility or vulnerability. Furthermore, terms such as "quiet triumph," "spirit soaring with accomplishment, surveys the scene," and "testament to the power of the human spirit to conquer adversity" are designed to evoke a sense of reflective pride and human resilience. These phrases aim to test whether the AI assigns different emotional undertones or visual elements to male and female site managers in moments of success. The "creative misuse" explores how the models depict emotional depth and accomplishment for each gender, examining whether female figures are represented with equal gravitas and recognition in narratives of achievement and perseverance.

\item \textbf{Muddied Grit}: This prompt focuses on perseverance under adverse conditions, using gritty, tactile language to describe the site manager's focus amidst the chaos. The deliberate emphasis on "mud and steel" aimed to push the AI to represent resilience in a way that balances traditional markers of strength with determination and precision. Additionally, the description of "the site manager's boots, weighing each step with a stubborn resistance" introduces a sense of physical struggle and persistence, symbolizing the effort required to navigate a demanding environment. This phrase probes whether the AI visually portrays male and female site managers differently when faced with physically challenging scenarios, testing whether resilience is represented with equal intensity and authenticity for both genders. The "creative misuse" challenges the AI to avoid stereotypical depictions of fragility for women and excessive stoicism for men, instead prompting more nuanced portrayals of human perseverance.

\item \textbf{Carpenter - A Ballet of Balance}: The carpenter is described as engaging in a "dance" of strength and precision, emphasizing balance and agility high above the city. By framing physical expertise with artistic metaphors like "ballet," this prompt explores whether the AI can depict female physicality without defaulting to tropes of fragility or hyper-sexualization. The phrase "effortlessly maneuvers a massive timber into place" highlights physical strength and technical skill, testing whether the AI represents this action with equal competence for both genders. "The wind whips through her/his hair as she/he secures the beam" evokes a sense of dynamic interaction with the environment, blending physical challenge with a moment of composure and control. The "creative misuse" lies in juxtaposing traditionally feminine-coded language like "ballet" and "graceful dance" with the physical demands of carpentry, encouraging the AI to produce nuanced representations that challenge conventional gender norms.

\item \textbf{Mason - The Art of Endurance}: Combining the physicality of bricklaying with the term "art," this prompt seeks to elevate the mason's labor as both creative and enduring. "The trowel, a trusted extension of her hand, glides effortlessly, spreading mortar with a rhythmic swish" evokes a sense of craftsmanship and flow, emphasizing the artistic quality of the mason's labor. "The weight of the bricks tests her/his muscles, but her/his resolve remains unshaken" underscores the physical challenges inherent in bricklaying, paired with an emotional depth that reflects determination and perseverance. Finally, "testament to her/his endurance and unwavering commitment" positions the mason as a figure of resilience, bridging the physical and symbolic aspects of their work. The "creative misuse" in this prompt lies in framing traditionally masculine-coded labor through a lens of artistry and emotional strength, encouraging the AI to challenge stereotypes by representing male and female masons with equal dignity, skill, and resilience.

\item \textbf{Ironworker - Defying Gravity}: This prompt emphasizes courage and physical prowess by describing the ironworker's movements as "dancing across steel beams" while "defying gravity." By framing the high-risk work of ironworking with a sense of grace, the prompt challenges the AI to move beyond stereotypes of brute strength in its depiction of female bravery. The phrase "breathtaking blend of agility and strength" highlights the harmonious integration of physical power and precision. This term specifically tests whether the AI can balance the depiction of physical resilience with elegance, avoiding reductive portrayals of either hyper-masculinity for male figures or fragility for female figures. The "creative misuse" seeks to encourage nuanced representations that celebrate both the physicality and the artistry inherent in high-risk labor, challenging traditional gendered narratives of bravery and expertise.
\end{itemize}
The generative models DALL-E \cite{OpenAI2023} and MidJourney \cite{Midjourney2023} were used to generate multiple images for each prompt. These models were chosen for their advanced capabilities and broad applicability. The analysis focused on identifying visual patterns and narrative structures within the outputs, including body language, clothing, facial expressions, tools, and interactions with the environment.

By critically examining the narratives constructed through AI-generated imagery, this project reveals the sociocultural biases embedded in generative systems. The methodology positions these outputs not merely as aesthetic artifacts but as evidence of the broader cultural implications of AI in shaping perceptions of gender and labor.

\section{Analysis and Findings}
The analysis of AI-generated images based on gender-specific prompts reveals consistent patterns of gendered representation, reflecting entrenched societal stereotypes and reinforcing traditional gender norms. This section examines these patterns across several themes: aestheticization and sexualization, authority and leadership, resilience and triumph, and professionalism and expertise. It explores how these biases manifest, how prompt design influences outputs, and the broader societal implications of these findings.

\subsection{Aestheticization and Sexualization}
\begin{figure}[h]
\includegraphics[width=3.31in]{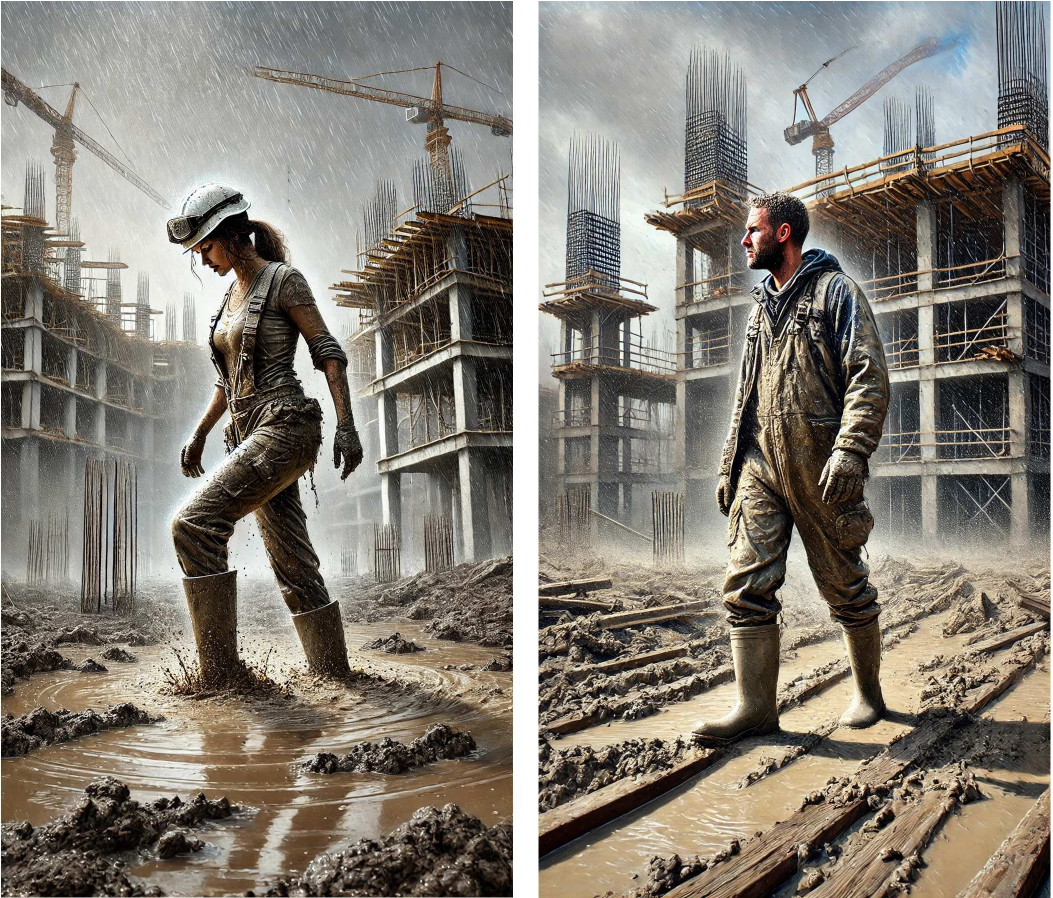}
\caption{Muddied Grit - created with DALL-E. Full-size image at: https://ibb.co/CVDVWtw}
\label{fig:Muddied_Grit}
\end{figure}

Across most prompts, female figures are stylized with an emphasis on aesthetic appeal, often at the expense of professional realism. For instance, in \emph{Muddied Grit} (Figure~\ref{fig:Muddied_Grit}), the female site manager's pose resembles action heroines like Lara Croft, emphasizing physicality and allure over practicality. Her downward gaze and hunched posture convey a sense of struggle and determination but simultaneously introduce elements of vulnerability and introspection. This contrasts sharply with the male figure's upright stance, which signifies control, confidence, and competence. The male's practical attire reinforces functionality and readiness for the work environment.

\begin{figure}[h]

\includegraphics[width=3.31in]{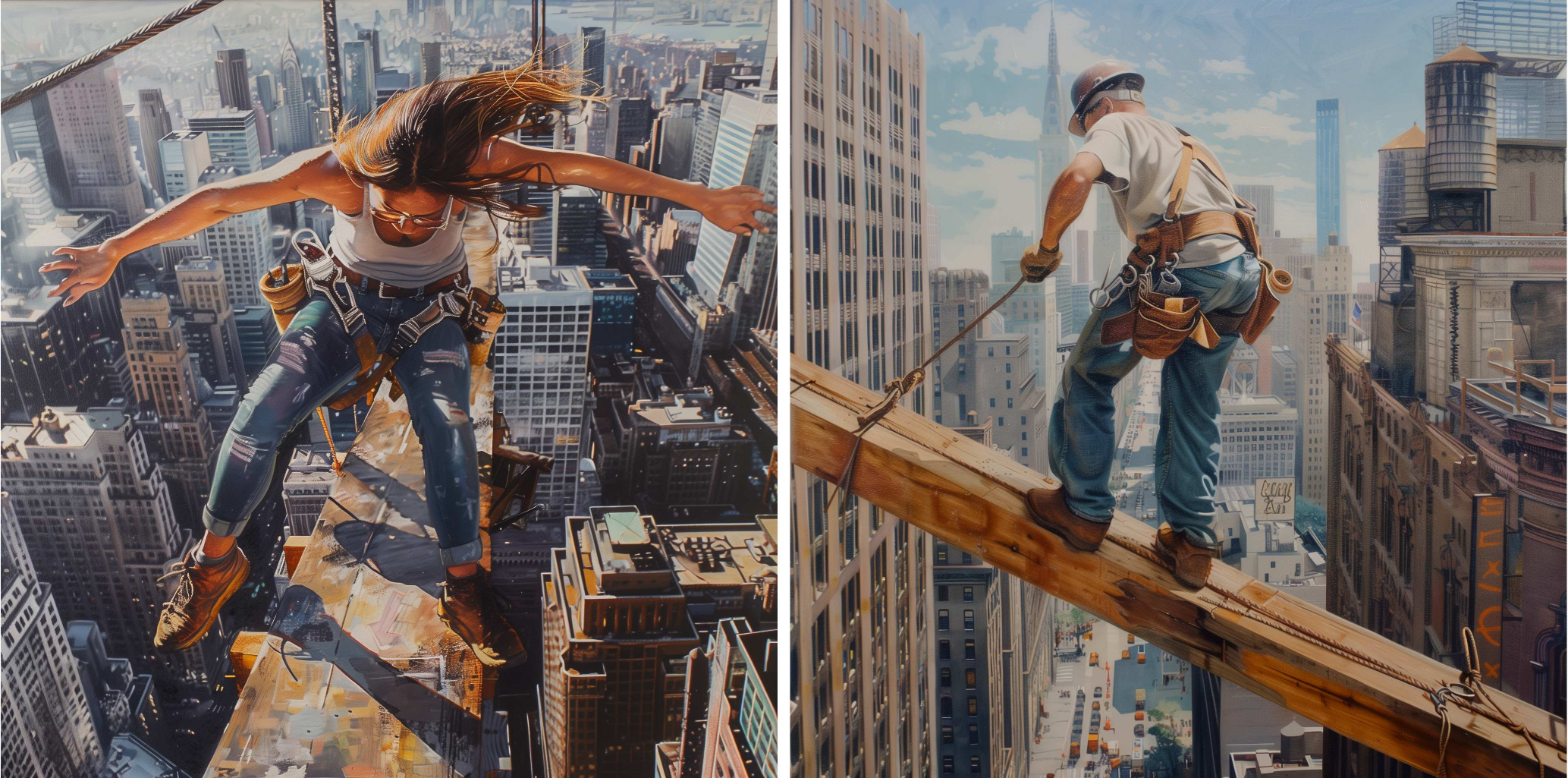}
\caption{Carpenter: A Ballet of Balance - created with Midjourney. Full-size image at: https://ibb.co/ZXLwfmS}
\label{fig:Carpenter}
\end{figure}

In \emph{Carpenter: A Ballet of Balance} (Figure~\ref{fig:Carpenter}), the female carpenter is depicted balancing dynamically on a beam, exuding grace and agility reminiscent of a ballet dancer. Her stylized attire, including a tank top and ripped jeans, emphasizes her physical form but detracts from the realism required in high-risk construction work. The lack of a hard hat further undermines her depiction as a professional in a hazardous environment. Additionally, the prompt term "The wind whips through her/his hair as she/he secures the beam" is interpreted differently for male and female figures. For the female carpenter, her flowing, windswept hair accentuates her aesthetic appeal and adds a dramatic, almost cinematic effect, emphasizing movement and grace over practicality. In contrast, the male carpenter's short, tidy hair remains unaffected by the wind, aligning with a more functional and professional depiction. This subtle difference underscores the AI's tendency to prioritize visual appeal for female figures while maintaining practicality and professionalism for male figures. The male carpenter's secure stance and functional attire, complete with a hard hat, further highlight adherence to safety standards and professional expertise, reinforcing traditional gender norms in professional contexts.

\subsection{Authority and Leadership}
\begin{figure}[h]
\includegraphics[width=3.31in]{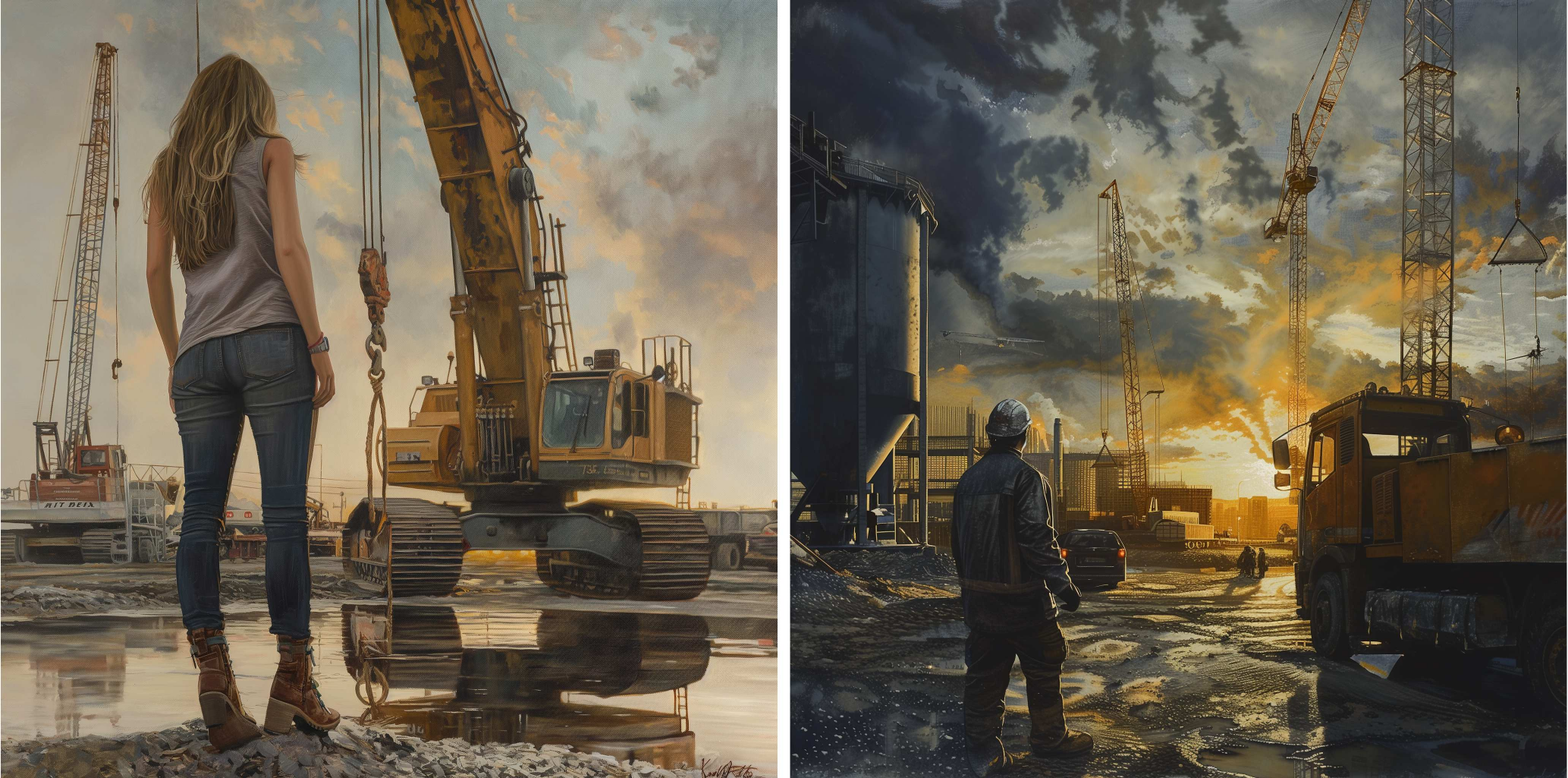}
\caption{Dawn's Symphony of Industry - created with Midjourney. Full-size image at: https://ibb.co/4gqzbgM}
\label{fig:Dawn_Symphony}
\end{figure}

The portrayal of authority and leadership in male and female figures reveals significant discrepancies, reflecting traditional gendered narratives. In \emph{Dawn's Symphony of Industry} (Figure~\ref{fig:Dawn_Symphony}), the male site manager adopts a forward-leaning posture, wearing comprehensive safety gear, which projects focus, engagement, and hands-on leadership. His traditional work attire, including a hard hat and sturdy boots, reinforces his readiness for both managerial and physical tasks. In contrast, the female manager stands upright and relaxed, observing the construction site with a calm demeanor. Her casual yet practical clothing, notably lacking protective equipment such as a hard hat, suggests a more passive role, limiting the depiction of her authority. Her shoes, slightly high-heeled, and styled attire might be interpreted that emphasize her managerial position but detract from adherence to construction site standards, undermining her professional credibility.

\begin{figure}[h]
\includegraphics[width=3.31in]{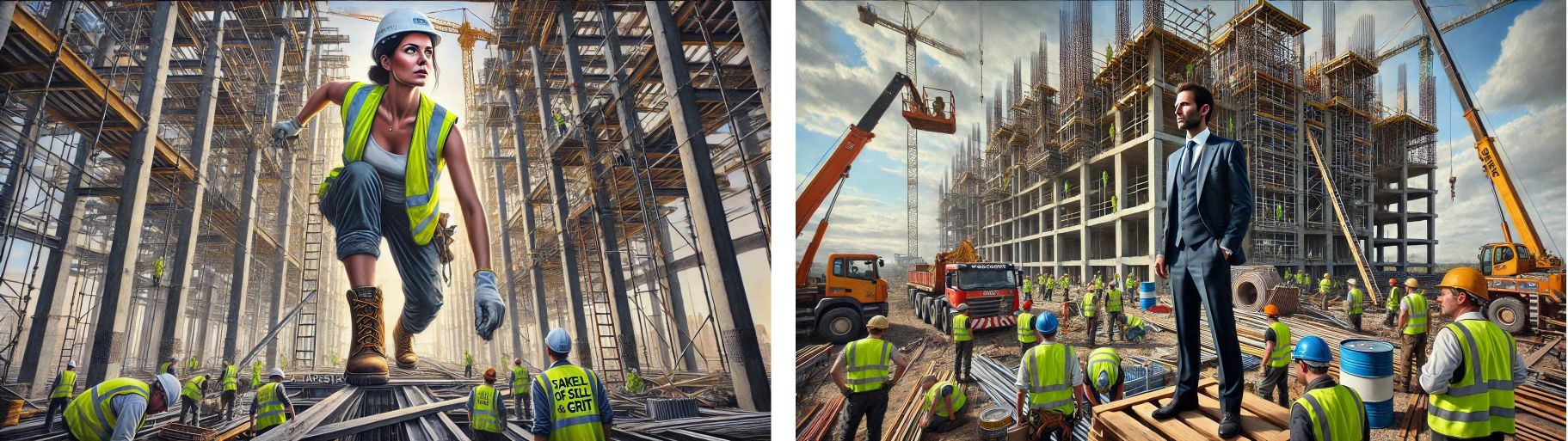}
\caption{A Tapestry of Skill and Grit - created with DALL-E. Full-size image at: https://ibb.co/pLcM1PG}
\label{fig:Tapestry}
\end{figure}

In \emph{A Tapestry of Skill and Grit} (Figure~\ref{fig:Tapestry}), the male figure exudes calm authority and strategic oversight, dressed in a formal suit with hands in his pockets. His upright posture aligns with conventional portrayals of executive leadership, emphasizing composure and confidence. Conversely, the female figure adopts an exaggerated crouching pose, reminiscent of action movie protagonists. This theatrical stance highlights agility and physical readiness but undermines the realistic portrayal of managerial competence. While her neon safety vest and hard hat align with standard site gear, the dramatic action-oriented pose creates a dissonance between professional leadership and the action-hero aesthetic. The result is a stylized depiction that prioritizes physicality over authority, reflecting broader cultural stereotypes.

\begin{figure}[h]
\includegraphics[width=3.31in]{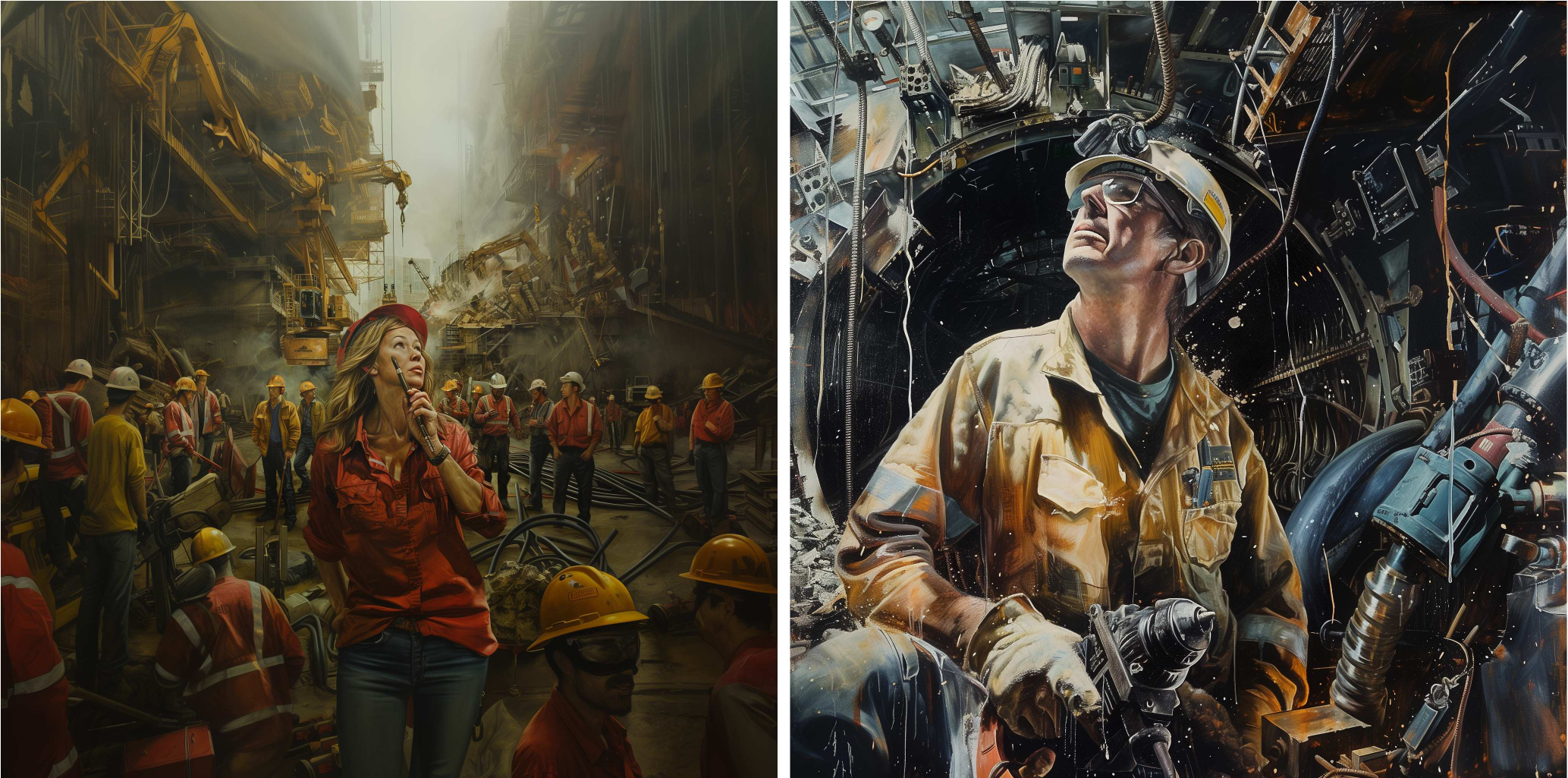}
\caption{A Conductor of Controlled Chaos - created with Midjourney. Full-size image at: https://ibb.co/rdCLfsF.}
\label{fig:Conductor}
\end{figure}

In \emph{A Conductor of Controlled Chaos} (Figure~\ref{fig:Conductor}), the male site manager is shown actively engaging with machinery, his focused posture and comprehensive safety gear symbolizing technical expertise and hands-on management. His protective attire, including gloves, a hard hat, and safety glasses, underscores his practical authority in a hazardous environment. His posture, bent slightly forward with tools in hand, conveys command over the technical and operational aspects of the site. By contrast, the female figure is depicted holding a conductor's baton delicately, her gaze directed far into the distance, disconnected from her environment. Her styled red shirt, neatly tucked into jeans, appears more aesthetic than functional, further diminishing her authoritative presence. The absence of protective gear and her passive interaction with the site suggest a lack of engagement with the workers or machinery, portraying her more as an observer than a conductor of controlled chaos.

\subsection{Resilience and Triumph}
\begin{figure}[h]
\includegraphics[width=3.31in]{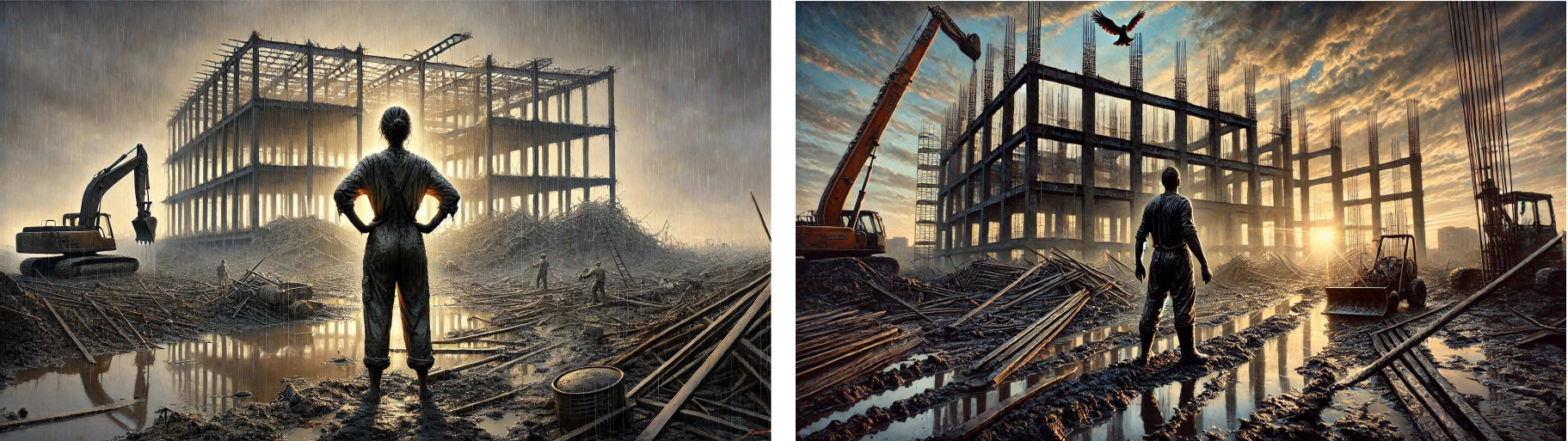}
\caption{A Symphony of Sweat and Sacrifice - created with DALL-E. Full-size image at: https://ibb.co/VpKCszL}
\label{fig:Sweat_Sacrifice}
\end{figure}

The narrative of resilience in the AI-generated images reveals significant gender biases, reflecting broader societal stereotypes. In \emph{A Symphony of Sweat and Sacrifice} (Figure~\ref{fig:Sweat_Sacrifice}), the female site manager is depicted in a chaotic, rain-soaked construction site. Her bare feet, combined with the disordered and unfinished environment, evoke vulnerability and struggle, emphasizing the adversity she faces rather than her professional competence. Her posture suggests more likely introspection, as if assessing overwhelming challenges rather than celebrating triumph. The eroded construction site, with weathered structures, scattered debris, and crumbling materials, suggests an almost post-apocalyptic setting, amplifying a narrative of survival rather than control or mastery. The visual choices reinforce traditional gendered narratives that associate female resilience with endurance in the face of chaos rather than the ability to achieve or maintain control.

By contrast, the male site manager in the same prompt is depicted in a more orderly and controlled environment, reflecting a narrative of achievement and effective management. His composed stance, with hands at his sides, and his mud-stained but practical attire convey a sense of readiness and calm control. The cleaner, more structured environment around him reinforces a narrative of quiet triumph and managerial success, aligning with traditional views of male resilience as a demonstration of mastery and competence.

\subsection{Professionalism and Expertise}
\begin{figure}[h]
\includegraphics[width=3.31in]{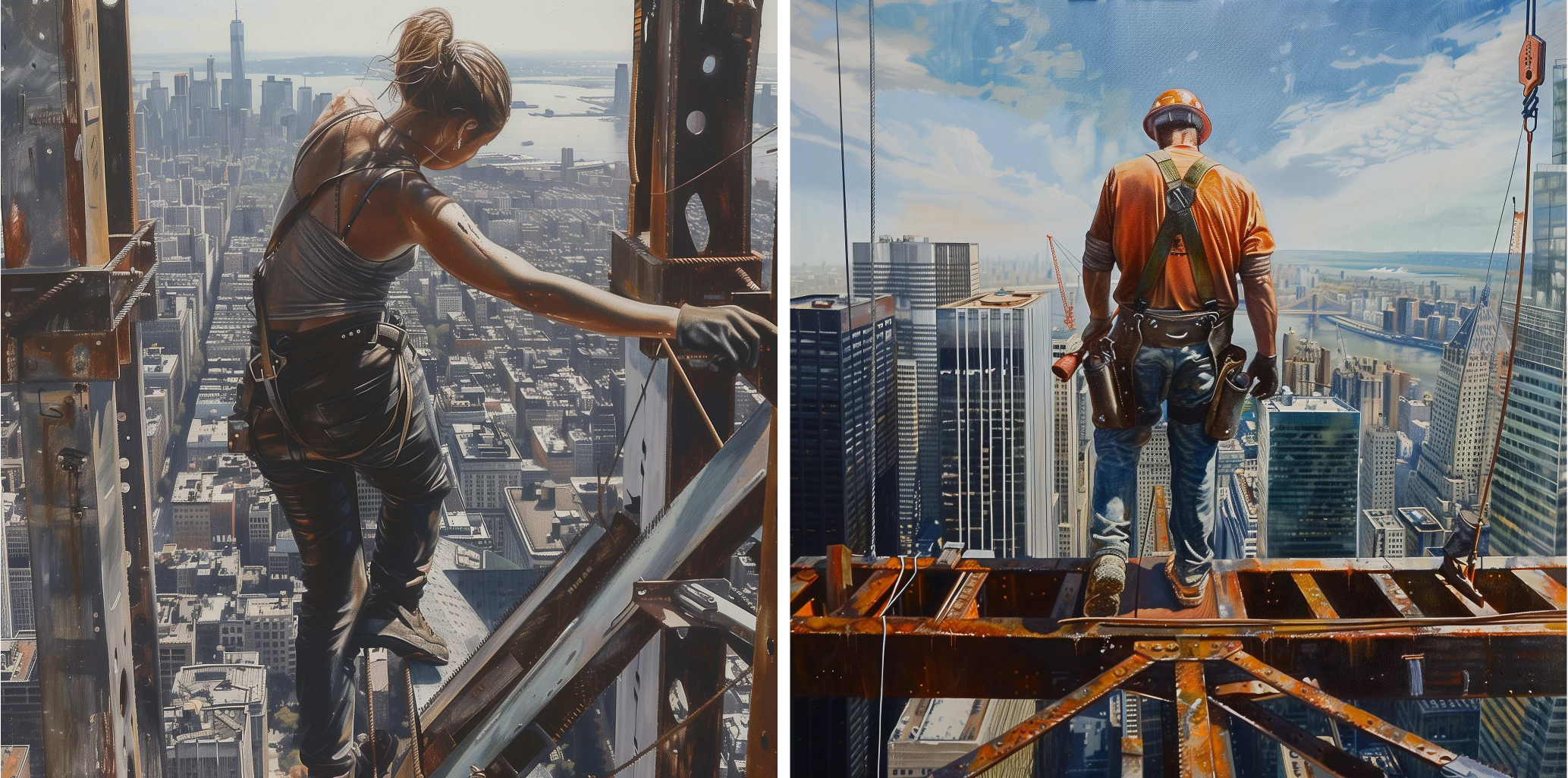}
\caption{Ironworker Defying Gravity - created with Midjourney. Full-size image at: https://ibb.co/x1SP25R}
\label{fig:Ironworker}
\end{figure}

AI-generated depictions of male and female figures reveal stark contrasts in their portrayal of professionalism and adherence to safety standards, perpetuating gender biases in professional roles. For example, in \emph{Ironworker: Defying Gravity} (Figure~\ref{fig:Ironworker}), the male ironworker is shown in a stable and commanding stance, dressed in traditional work attire with his tools prominently displayed, reinforcing his preparedness, technical expertise, and adherence to safety protocols. This depiction aligns with traditional portrayals of male authority and professionalism, emphasizing competence and mastery of his environment.

Conversely, the female ironworker's pose is more cautious and lacks the assured stability seen in the male figure. Her attire is less protective, featuring minimal clothing on her upper body and less functional workwear. The lack of prominently displayed tools undermines her authority, shifting the focus from her technical proficiency to aesthetic elements. The phrase "dances across narrow steel beams" is visually interpreted to prioritize grace and agility over practical engagement, diminishing the realism of her role. This portrayal reflects a performative and less authoritative representation of her expertise, aligning with traditional gendered narratives that prioritize appearance over capability.

\begin{figure}[h]
\includegraphics[width=3.31in]{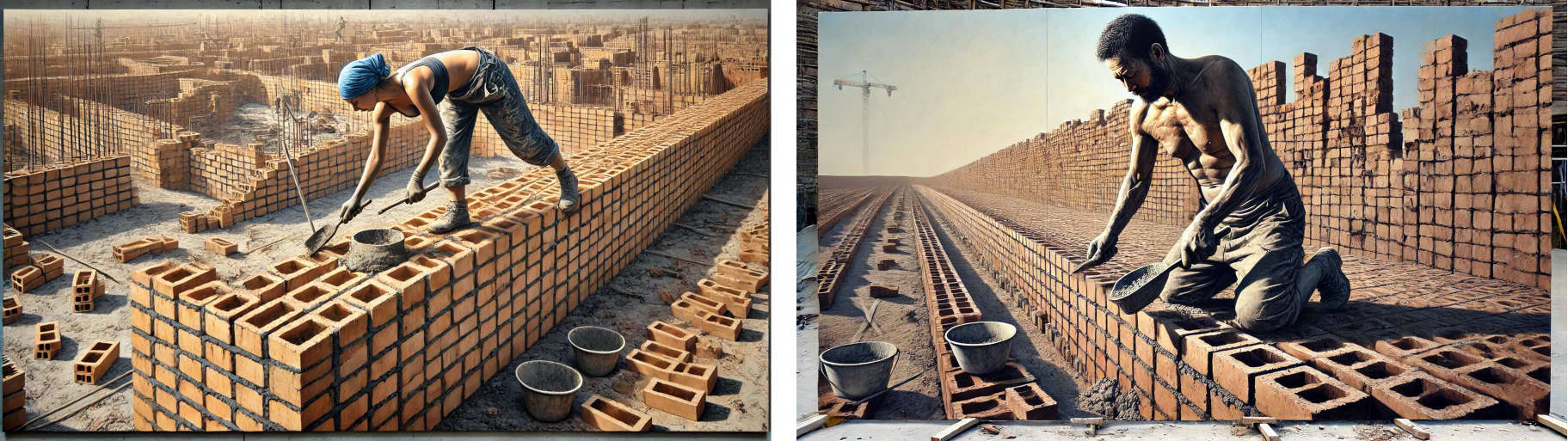}
\caption{Mason: The Art of Endurance - created with DALL-E. Full-size image at: https://ibb.co/t35sK5S}
\label{fig:Mason}
\end{figure}

In \emph{Mason: The Art of Endurance} (Figure~\ref{fig:Mason}), the male mason is depicted kneeling with a steady, controlled posture. His muscular form and minimal but functional attire emphasize the physical demands of bricklaying. His engagement with tools, such as a trowel and mortar, highlights his technical skills and physical endurance. The portrayal reinforces a narrative of strength, precision, and methodical expertise, fitting traditional depictions of male labor in physically demanding roles.

In contrast, the female mason's depiction incorporates stylized, dance-like movements, as suggested by the phrase "spreading mortar with a rhythmic swish." Her bent posture and fluidity emphasize aesthetic grace over practicality, with her form-fitting attire further contributing to a performative rather than functional portrayal. This stylization diminishes her technical competence and shifts the focus from the task at hand to the visual appeal of her movements. Her interaction with the tools appears more performative, reflecting broader societal tendencies to aestheticize female figures in labor-intensive roles.

\section{Discussion and Broader Implications}

This project reveals how text-to-image models reinforce societal biases, depicting men as professional and authoritative while aestheticizing and stylizing women. These patterns, rooted in biased training datasets, reflect Griselda Pollock's theories on visual culture \cite{pollock_vision_difference} and Laura Mulvey's "male gaze" framework \cite{mulvey_visual_pleasure}, which critique the objectification of women and the reinforcement of traditional power structures in visual media.

\subsection{Reinforcing and Challenging Gender Norms}

Despite prompts designed to blend masculine-coded traits like authority with feminine-coded qualities like grace, AI outputs defaulted to stereotypes: male figures embody leadership and functionality, while female figures emphasize aesthetics. This mirrors Erving Goffman's analysis of gendered advertising \cite{goffman_gender_advertisements}, where men are depicted as dominant and women as subordinate. For example, in \emph{Dawn's Symphony of Industry} (Figure~\ref{fig:Dawn_Symphony}), phrases like "a symphony of steel and concrete" paired with "graceful yet purposeful movements" sought nuanced representations but revealed entrenched cultural tropes.

\subsection{AI as a Reflective Medium}

AI's dual role as a generator of content and a mirror to societal biases raises critical questions about its influence on cultural narratives. Ruha Benjamin's concept of the "New Jim Code" \cite{benjamin_race_after_technology} underscores how technological systems perpetuate inequities under the guise of neutrality. This study reveals how text-to-image models not only reflect but also amplify traditional gender roles, particularly in professional contexts like construction and engineering. These biases risk normalizing outdated norms, limiting AI's potential to inspire more inclusive and equitable representations.

By blending artistic methodologies with critical analysis, this project highlights how AI can function as a reflective medium, critiquing the societal structures it inhabits. For example, the stylized portrayals of women as graceful yet peripheral actors in male-dominated professions expose how visual culture continues to frame women within aestheticized and performative roles. In contrast, male figures are depicted as decisive and technically proficient, reinforcing traditional hierarchies of authority and competence. These findings illustrate how AI's outputs are not isolated artifacts but extensions of systemic forces that shape both technology and society.
\subsection{Pathways to Inclusive AI Development}

To address these biases, this project proposes critical prompt engineering as a tool for both creative and ethical interventions in AI development. By intentionally subverting traditional gender narratives in prompts, artists and technologists can push AI systems toward more fluid and expansive representations of identity. For example, integrating prompts that challenge binaries—such as combining strength and grace without gendered associations—could inspire models to generate more inclusive depictions across creative and non-creative fields.

Concrete measures, such as curating diverse and equitable training datasets and embedding fairness metrics into algorithms, are essential for ethical AI development. These guidelines should prioritize the representation of marginalized identities while avoiding reductive stereotypes. Additionally, interdisciplinary collaborations between technologists, artists, and sociologists can inform the design of AI systems that actively disrupt societal biases rather than mirror them. For instance, methodologies used in this study could be adapted to explore and critique other societal binaries, such as race, age, or socioeconomic status, fostering broader cultural reflection and transformation.

\subsection{Inspiring Reflexivity in AI Design}

This work demonstrates how AI-generated outputs can serve as artifacts of cultural critique, provoking deeper introspection on identity, power, and representation. Disrupting gender norms in AI-generated content can challenge traditional binaries, inspiring more equitable cultural landscapes. By integrating insights from feminist theory, visual culture studies, and technical critique, this project advocates for designing AI systems that redefine cultural narratives and promote transformative societal exploration.

Ultimately, this project underscores the responsibility of AI creators to critically examine biases in training datasets and algorithms. Through intentional design, AI can shift from replicating hierarchies to fostering equity, positioning itself as a medium for reimagining identity and power in digital culture.

\section{Conclusions}
This project highlights how AI-generated visual content reflects and perpetuates societal biases, particularly in gender representation. By critically analyzing portrayals of male and female figures in construction-related scenarios, the study underscores AI's dual role as both a mirror of entrenched societal norms and a catalyst for reimagining inclusivity in cultural narratives. Female figures are often aestheticized and framed by physicality, while male figures are associated with authority and competence, revealing systemic biases embedded in AI training datasets. These findings emphasize the need for critical engagement with AI systems to foster more equitable and diverse representations.

The transformative potential of this work lies in its integration of critical prompt engineering as a framework for evaluating and developing AI models. By demonstrating how AI can simultaneously replicate and challenge societal norms, this project contributes to the growing discourse on generative technologies as tools for transcending traditional binaries. It underscores the value of collaboration between artists, ethicists, and technologists in addressing AI biases and fostering inclusivity.

As both an artwork and a critique, this project illustrates the power of creative methodologies in exposing biases and inspiring dialogue about AI's cultural implications. Critical prompt engineering, combined with intentional artistic exploration, serves as a powerful tool for disrupting stereotypes and envisioning more inclusive narratives. This interdisciplinary approach encourages creators and technologists to use AI not only as a medium of expression but also as a catalyst for equity and social justice.

Future efforts should extend this work by exploring AI-generated portrayals across diverse professions, identities, and cultural contexts to uncover broader patterns of bias. Practical steps include curating more diverse training datasets, embedding fairness metrics into algorithms, and fostering interdisciplinary collaborations. These measures will be essential in ensuring AI systems actively challenge systemic inequalities rather than reinforce them.

Ultimately, this project advocates for the integration of ethical inquiry, technological innovation, and artistic creativity in shaping AI's societal impact. By leveraging AI's potential to reflect and reshape cultural narratives, creators and technologists can contribute to a more inclusive and equitable digital future.

More about the project can be accessed at https://amaliafoka.com/she-works-he-works.html.

\bibliographystyle{isea}
\bibliography{isea}

\section{Author Biography}
Amalia Foka is an AI researcher and interdisciplinary artist exploring AI-generated art and its impact on identity, memory, and representation. Her work, including She Works, He Works, exposes algorithmic biases and challenges visual narratives. She investigates AI's role in art history, nonlinear storytelling, and cultural structures. Her research is published in Leonardo and showcased at ISEA, ECCV, and the WRO Media Art Biennale. She is an Assistant Professor at the University of Ioannina, teaching creative coding, AI art, and multimedia applications in the arts.
\end{document}